\documentstyle[twoside,fleqn,espcrc2,epsf]{article}

\newcommand{\lw}[1]{\smash{\lower2.ex\hbox{#1}}}

\newcommand{\AmS}{{\protect\the\textfont2
  A\kern-.1667em\lower.5ex\hbox{M}\kern-.125emS}}

\title{
\vspace*{-32pt}
{\normalsize \hfill UTHEP-372} \\
\vspace*{-5pt}
{\normalsize \hfill UTCCP-P-27} \\
\vspace*{-5pt}
{\normalsize \hfill October 1997} \\
CP-PACS results for quenched QCD spectrum with the Wilson action
\thanks{Talk presented by T.\ Yoshi\'e at the International Workshop on 
``LATTICE QCD ON PARALLEL COMPUTERS'', 10-15 March 1997, Center for
Computational Physics, University of Tsukuba.}
}

\author{CP-PACS Collaboration \\[2mm]
        S.~Aoki\address{Institute of Physics, University of
        Tsukuba, Tsukuba, Ibaraki 305, Japan},
	G.~Boyd\address{Center for Computational Physics,
        University of Tsukuba, Tsukuba, Ibaraki 305, Japan},
	R.~Burkhalter$^{\rm b}$, 
        S.~Hashimoto\address{High Energy Accelerator Research Organization
        (KEK), Tsukuba, Ibaraki 305, Japan},
        N.~Ishizuka$^{\rm a}$,
        Y.~Iwasaki$^{\rm a,b}$,
        K.~Kanaya$^{\rm a,b}$,
        Y.~Kuramashi$^{\rm c}$,
        M.~Okawa$^{\rm c}$, A.~Ukawa$^{\rm a}$,
        T.~Yoshi\'e$^{\rm a,b}$ }

\begin{document}

\begin{abstract}
We present progress report of a CP-PACS calculation of
quenched QCD spectrum with the Wilson quark action.
Light hadron masses and meson decay constants are obtained
at $\beta=$5.9,\ 6.1, and 6.25 on lattices with a physical 
extent of 3 fm, and for the range of quark mass
corresponding to $m_\pi/m_\rho \approx 0.75$ $-$ 0.4. 
Nucleon mass at each $\beta$ appears to be a convex function
of quark mass, and consequently the value at the physical quark 
mass is much smaller than previously thought.
Hadron masses extrapolated to the continuum limit exhibits a significant 
deviation from experimental values: with $K$ meson mass to fix strange quark 
mass, strange meson and baryon masses are systematically lower. 
Light quark masses determined from the axial Ward identity 
are shown to agree with those from perturbation theory in the continuum
limit.  Decay constants of mesons are also discussed.
\vspace{-0.1cm}
\end{abstract}

\maketitle

\section{Introduction}

Precision determination of the light hadron spectrum is a fundamental
problem of lattice QCD.  Over the years numerous studies have been carried out 
on this subject, which have revealed a number of difficulties that have 
to be overcome to achieve this goal.  In addition to the overriding issue of 
how quenching distorts the spectrum, 
calculations aiming at precise results have to have good control 
over the systematic errors arising 
from a finite spatial lattice size, an extrapolation toward light 
quark masses, and an extrapolation to the continuum limit.
There is also the issue of how reliably one can extract 
hadron masses from the propagators whose statistical errors generally grow 
exponentially with the distance from the source point.

Within quenched QCD high statistics calculations on large lattices attempting 
to deal with these problems gradually developed since around 1990 
exploiting the power of 
dedicated parallel 
computers\cite{ref:APE24,ref:QCDPAX96,ref:GF11mass,ref:LANL96,ref:MILC97}.  
In particular 
Weingarten and collaborators with the GF11 computer carried out a pioneering 
work on the continuum limit of the quenched spectrum with extensive 
simulations including finite size studies\cite{ref:GF11mass}. 
In spite of these efforts convincing results on hadron masses in quenched QCD
with a precision reliably better than 5\% have not been obtained.  
In particular 
the question in what way the quenched spectrum deviates from the experiment
has not been fully answered so far.

In view of this situation the CP-PACS Collaboration has decided to undertake 
a quenched simulation of the light hadron spectrum as the first lattice 
QCD project of the CP-PACS computer which started operation in April 1996.
The strategy we have chosen is to employ the simplest form of the action, 
{\it i.e.,} the plaquette action for gluons and the Wilson action for quarks, 
and carry out, as much as the computing power of CP-PACS allows, 
a measurement of hadron masses down to small quark masses at 
small lattice spacings employing large lattices and high statistics.
The simulation has been running since the summer of 1996 and are still 
continuing.
In this article we report the present status of the work.

\section{Parameters of simulation}

\begin{table*}[t]
\caption{Simulation parameters.
Values of $a^{-1}$ are determined from $m_\rho$.}
\label{tab:parameter}
\begin{center}
\begin{tabular}{lcccccc}
\hline\hline
\lw{$\beta$} & \lw{size}  &  $a^{-1}$  & $La$ & \lw{\#conf} & 
sweep  & \lw{hopping parameter} \\
          &               &  (GeV)     & (fm) & &
/conf  & \\
\hline
5.90\ \   & $32^3\times \,\, 56$  & 1.96(\, 2)    & 3.23(3)   & 674 & 
200 & 0.15660,\ 0.15740,\ 0.15830,\ 0.15890,\ 0.15920 \\
6.10\ \   & $40^3\times \,\, 70$  & 2.58(\, 3)    & 3.06(3)   & 384 & 
400 & 0.15280,\ 0.15340,\ 0.15400,\ 0.15440,\ 0.15460 \\
6.25\ \   & $48^3\times \,\, 84$  & 3.10(\, 4)    & 3.05(4)   & 262 & 
1000 & 0.15075,\ 0.15115,\ 0.15165,\ 0.15200,\ 0.15220 \\
6.47\ \   & $64^3\times 112$ & 4.19(12)      & 3.01(8)   & 22  & 
2000 & 0.14855,\ 0.14885,\ 0.14925,\ 0.14945,\ 0.14960 \\
\hline\hline
\end{tabular}
\end{center}
\vspace{-8mm}
\end{table*}

In Table \ref{tab:parameter} we list the parameters chosen for our simulation,
and the number of configurations for hadron mass measurements accumulated 
as of the time of the Workshop.

We choose four values of the coupling constant $\beta$ to cover 
the range of lattice spacing $a^{-1}\approx 2-4$ GeV 
for controlling scaling violation effects and taking the continuum limit.
In order to avoid finite size effects, the spatial lattice size $L$ is 
chosen so that the physical size equals approximately 3 fm.  
The temporal lattice size is set equal to $7L/4$; this choice is based on 
considerations on length of time intervals needed for a reliable 
extraction of masses with an exponentially smeared source we employ 
for hadron mass measurements.  

For each value of $\beta$ we select five values of the hopping parameter 
such that the ratio $m_\pi/m_\rho$ takes values of about 
0.75,\ 0.7,\ 0.6,\ 0.5, and 0.4.  We abbreviate these hopping parameters 
as $s_1$,\ $s_2$,\ $u_1$,\ $u_2$ and $u_3$.
The first two values are for interpolation of results 
to strange quark, and the rest are for examining chiral extrapolation.  
Previous spectrum studies with the Wilson quark action have been limited 
to the range  $m_\pi/m_\rho\geq 0.5$.  The value $m_\pi/m_\rho\approx 0.4$
represents our attempt toward lighter quark masses. Reducing the quark mass 
further is not easy: test runs we have carried out for 
$m_\pi/m_\rho\approx 0.3$ at $\beta=5.9$ show that fluctuations large and 
that it takes more computer time than that for the 
five hopping parameters down to  $m_\pi/m_\rho\approx 0.4$.
We calculate hadron masses for equal mass cases and also for the unequal 
mass cases of the type $s_i u_j$ for mesons and
$s_i s_i u_j$ and $s_i u_j u_j$ for baryons.

Configurations are generated with the 5-hit Cabibbo-Marinari-Okawa
heat-bath algorithm and the over-relaxation algorithm 
mixed in the ratio of 1:4.  The combination is called a sweep, and we 
skip 200 to 2000 sweeps between hadron mass measurements as listed in 
Table \ref{tab:parameter}.  

Quark propagators are solved with the red/black-preconditioned minimal 
residual algorithm imposing the periodic boundary condition in all 
four directions. The stopping condition is chosen to ensure that
truncation error in hadron propagators is at most 5 percent of
our estimated final statistical error.
For performance of our code on the CP-PACS see 
Refs.~\cite{ref:IwasakiCCPWS,ref:GordonBell} 

Errors are estimated by a single elimination jackknife procedure.

We note that results presented here are preliminary,
as runs are still continuing.  In particular the run at $\beta=6.47$ 
has been just started. 
Although some figures include results at $\beta=6.47$,
values in the continuum limit are determined
by extrapolating data at the first three values of $\beta$. 

\section{Extraction of hadron masses}
\subsection{Smearing}\label{sec:smearing}

We employ an exponential smearing of quark source to enhance signals of 
ground state in hadron propagators.  
This choice is motivated by a recent measurement of the pion wave function 
defined by 
$\psi(r) = \langle 0|\sum_x \bar q(x) \gamma_5 q(x+r)|\pi\rangle
/\langle 0|\sum_x \bar q(x)\gamma_5 q(x)|\pi\rangle$ by the 
JLQCD Collaboration, who found that $\psi(r)$ is well reproduced by a single
exponential function: $\psi (r)=A \exp( -B r)$\cite{ref:HashimotoLAT96}.

To choose the value of the slope $B$ in our measurement, 
we parameterize their results in the form 
$B(m_\rho,m_q) =  (x_0 + x_1 \cdot m_\rho) +
(y_0 + y_1 \cdot (m_\rho)^2)\cdot m_q$.  
Estimates from available data for the value of $m_\rho$ 
appropriate for our simulation parameters are then substituted to find the
value of $B$. 
The smearing radius $1/B$ chosen in this way does not vary much over 
the points of our simulation, 
being approximately given by $1/B\approx 0.33$ fm.
  
\noindent
\begin{figure}[t]
\begin{center} \leavevmode
\epsfxsize=7.5cm \epsfbox{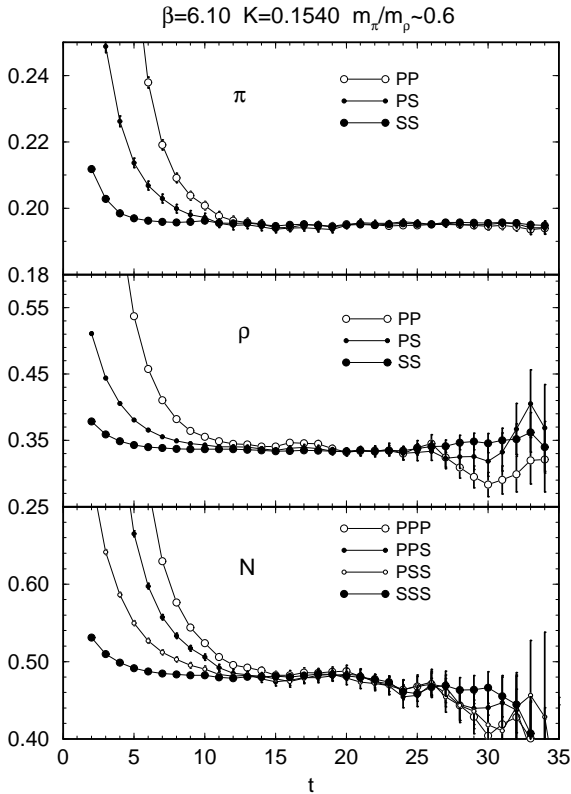}
\end{center}
\vspace{-12mm}
\caption{Typical results of effective masses
obtained with various combinations of quark sources.}
\label{fig:smear}
\vspace{-6mm}
\end{figure}
Various combinations of point (P) and smeared (S)
sources are used to construct hadron operators at the source.
In the following we employ a notation such as PSS to specify the combination 
of quark sources. On the other hand, we always use point sinks for 
hadron operators at the sink.

In Fig.\ref{fig:smear} we show typical results of effective masses for various 
source combinations.
We find that the effective mass reaches a plateau from above 
in almost all cases for all combinations of source.  This suggests that 
the smearing radius of $1/B\approx 0.33$ fm
we choose is smaller than the actual spread of hadron wave functions.

\subsection{Fitting Procedure}\label{sec:fitting}

We observe in Fig.~\ref{fig:smear} that the onset of plateau is earliest 
if the smeared source is used for all quark and antiquark fields in hadron 
operators.  Furthermore statistical errors of effective masses 
at a given time slice are the smallest for this combination.
For results presented in this report, we therefore decide to 
derive masses from the SS source for mesons and the SSS source for 
baryons.

To extract hadron masses from propagators we employ a single hyperbolic 
cosine fit for mesons and a single exponential fit for baryons. 
The largest time slice of the fit $t_{\rm max}$ is chosen by the 
requirement that the error of propagator does not exceed 5\%.
Changing the minimum time $t_{\rm min}$ of the fit, we compare results of 
correlated and uncorrelated fits.
We find that 
1) $\chi^2/N_{DF}$ for correlated fit decreases as $t_{\rm min}$ increases
and becomes constant at some time slice called $t_\chi$\cite{ref:QCDPAX96}.
2) For $t_{\rm min} \ge t_\chi$,
mass results from correlated and uncorrelated fits agree 
within at most 1.5 standard deviations.
3) Uncorrelated fits give results more stable as a function of $t_{\rm min}$.
From these investigations, we decide to adopt mass results from the
uncorrelated fit at $t_{min}\approx t_\chi$, supplememted with 
a single elimination jackknife analysis to estimate the error of masses.
We consider that the magnitude of systematic error due to the choice 
of fitting ranges and fitting procedure is comparable to statistical error.

In Table \ref{tab:error}, we list typical statistical errors in mass results
in units of percent, which are estimated by the single elimination
jackknife method. In the contiuum limit results for the nucleon has the 
largest error of about 5\% with present statistics.

\begin{table}[t]
\caption{Errors in mass results in units of percent.}
\label{tab:error}
\vspace{-2mm}
\begin{center}
\begin{tabular}{lcccc}
\hline\hline
   &  $\pi$ & $\rho$ & $N$ & $\Delta$ \\
$s_1$ & 0.1 & 0.2 & 0.3 & 0.3 \\
$s_2$ & 0.1 & 0.3 & 0.3 & 0.4 \\
$u_1$ & 0.2 & 0.4 & 0.5 & 0.5 \\
$u_2$ & 0.3 & 0.6 & 0.7 & 0.6 \\
$u_3$ & 0.4 & 0.8 & 0.9 & 0.7 \\
$K_c$ &     & 1.2 & 2.0 & 1.0 \\
$a=0$ &     &     & 5.5 & 3.7 \\
\hline\hline
\end{tabular}
\end{center}
\vspace{-8mm}
\end{table}

\section{Extrapolation and interpolation to physical quark mass}
\label{sec:extrap}

\noindent
\begin{figure}[t]
\begin{center} \leavevmode
\epsfxsize=7.5cm \epsfbox{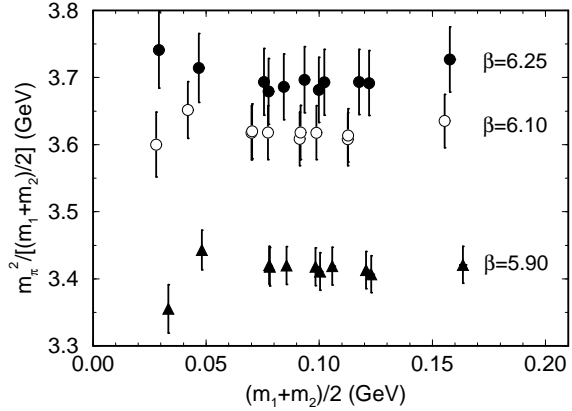}
\end{center}
\vspace{-12mm}
\caption{$m_\pi^2/[(m_1+m_2)/2]$ as a function of $(m_1+m_2)/2$.}
\label{fig:pi2mq}
\vspace{-6mm}
\end{figure}
Our first interest is how hadron masses depend on the quark mass.
We start with the pseudo scalar case, and plot in Fig.\ref{fig:pi2mq} 
the ratio 
$m_\pi^2/[(m_1+m_2)/2]$ against $(m_1+m_2)/2$, 
where $m_i$ are quark masses defined by $m_i = Z_m(1/K_i-1/K_c)$. Conversion 
to physical units is made with estimates of lattice spacing in terms of 
$m_\rho$ as described below.  
The critical hopping parameter $K_c$ is determined by a linear 
extrapolation of $m_\pi^2$, and $Z_m$ denotes the one-loop renormalization 
factor (see sec.\ref{sec:quarkmass} for details). 
The ratio for 11 combinations of quark masses is constant within errors 
at each $\beta$.
Therefore we conclude that the pseudo scalar meson mass squared is
linear in the average quark mass in the range covered by our data.
In other words, we find no clear evidence for the existence of 
chiral perturbation theory higher order terms 
or quenched chiral logarithms\cite{ref:SBGXLOG} in the results 
of pseudo scalar meson masses down to $m_q \approx$ 30 MeV.

\begin{figure}[t]
\begin{center} \leavevmode
\epsfxsize=7.5cm \epsfbox{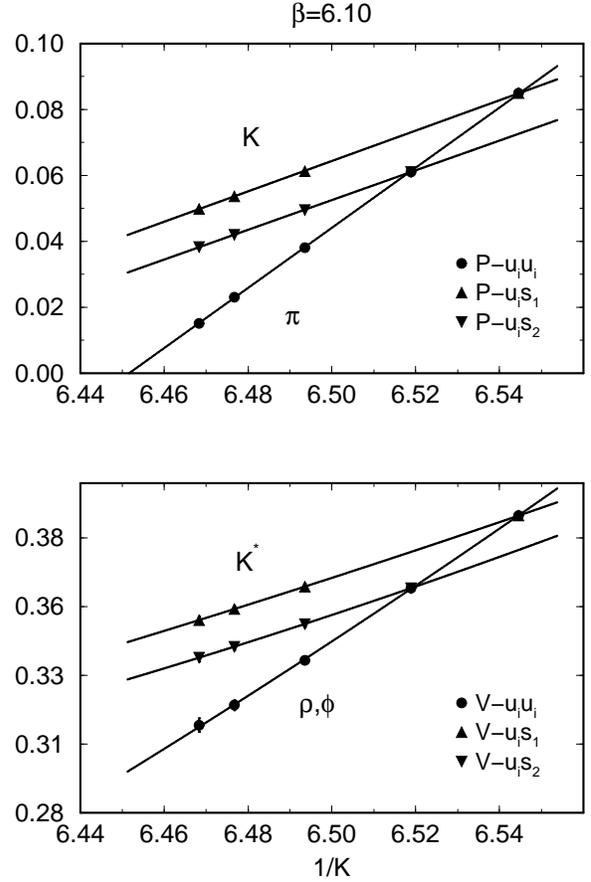}
\end{center}
\vspace{-12mm}
\caption{Pseudo-scalar meson masses squared
and vector meson masses as functions of $1/K$ of light quark.}
\label{fig:meson}
\vspace{-6mm}
\end{figure}

\begin{figure}[t]
\begin{center} \leavevmode
\epsfxsize=7.5cm \epsfbox{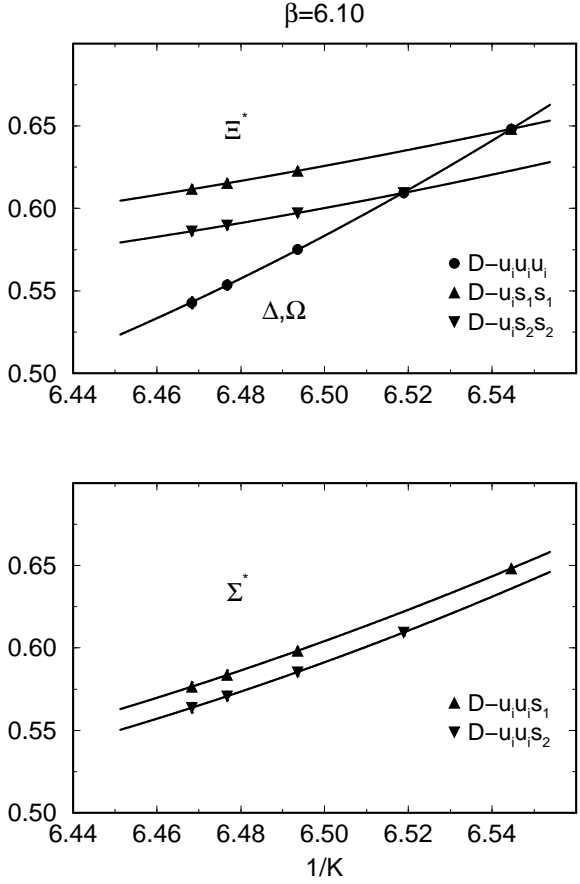}
\end{center}
\vspace{-12mm}
\caption{Decuplet baryon masses against $1/K$ of light quark.}
\label{fig:decuplet}
\vspace{-6mm}
\end{figure}

\begin{figure}[t]
\begin{center} \leavevmode
\epsfxsize=7.5cm \epsfbox{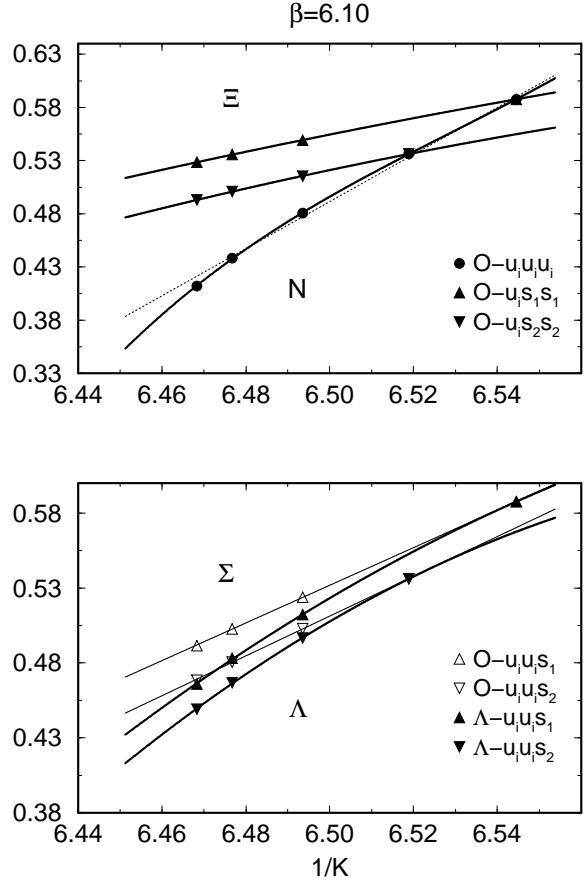}
\end{center}
\vspace{-12mm}
\caption{Octet baryon masses against $1/K$ of light quark.}
\label{fig:octet}
\vspace{-6mm}
\end{figure}

\begin{figure}[t]
\begin{center} \leavevmode
\epsfxsize=7.5cm \epsfbox{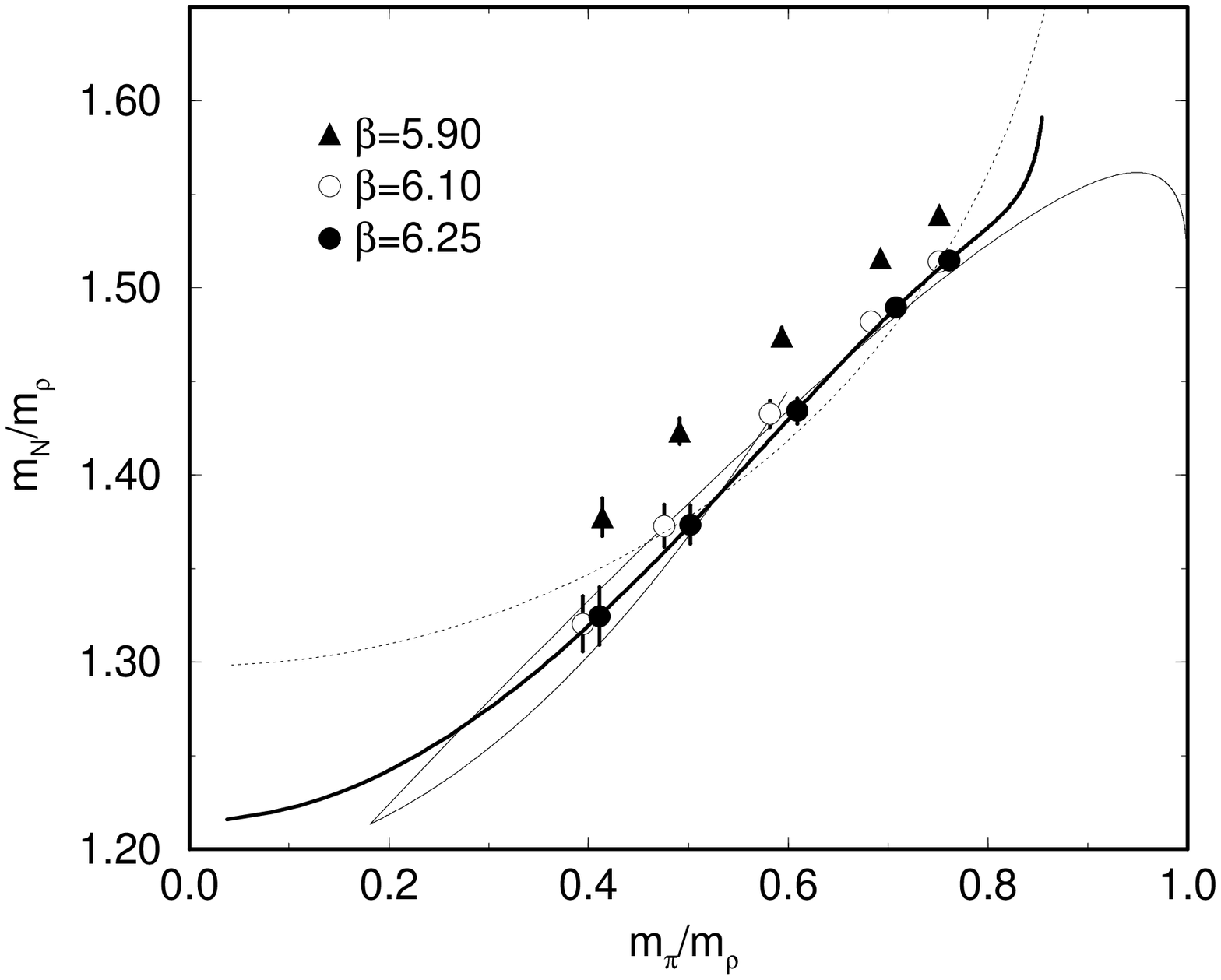}
\end{center}
\vspace{-12mm}
\caption{Edinburgh plot showing $m_N/m_\rho$ as a function of
$m_\pi/m_\rho$. For explanation of lines see text.}
\label{fig:edplot}
\vspace{-6mm}
\end{figure}

Vector mesons and decuplet baryons also exhibit the property that their 
masses are 
linear in the average of quark masses with only a small mixture of 
higher order terms.  This is illustrated in Fig.\ref{fig:meson} 
and \ref{fig:decuplet} in which hadron masses are plotted
against $1/K$ of lighter quark, together with results of a linear 
(pseudo scalar case) or quadratic (vector and decuplet case) fit.

However,  this property is badly broken for octet baryons as shown in 
Fig.\ref{fig:octet}.
Although $\Sigma$ and $\Xi$ masses are linear in $1/K$,
nucleon and $\Lambda$ masses are convex. 
The nucleon mass at the lightest quark mass is clearly off the linear 
fit (dashed line) as compared to a cubic fit shown by a solid line. 

In this connection we show our results for $m_N/m_\rho$ as a function of 
$m_\pi/m_\rho$ in Fig.\ref{fig:edplot}.
Dotted and thick curves correspond to the cases in which 
the nucleon mass at $\beta =$ 6.25 is fitted with a linear
or cubic functions of quark mass, respectively.
Higher order terms are essential in order to make a reliable extrapolation 
of nucleon and $\Lambda$ masses to the chiral limit, 
and have a significant impact 
on final results.

Based on the considerations described above, we employ a linear fit 
for pseudo-scalar mesons,
cubic fit for nucleon and quadratic fit for other hadrons 
in order to extrapolate or interpolate hadron mass results
to physical quark masses at each $\beta$.
For vector mesons a linear fit reads to essentially the same result.
We then determine values of $K_{u,d}$ and $a^{-1}$ taking 
the experimental values of $m_\pi$ and $m_\rho$.  
The hopping parameter $K_s$ for strange quark is estimated in two ways, 
either from $m_\phi$ assumed to be a pure $\overline{s}s$ state or from 
$m_K$ or $m_{K^*}$.  For the latter case we take 
meson mass data for unequal quark masses and first 
make an extrapolation of lighter quark to $K_{u,d}$ with the heavier 
quark fixed at $s_1$ and $s_2$.  
We then linearly interpolate the result so that $K$ or $K^*$ meson mass 
has the experimental value.
For experimental values we use $m_\pi=135$ MeV, $m_\rho=769$ MeV,
$m_\phi=1019$ MeV, $m_K=498$ MeV, and $m_{K^*}=896$ MeV.

\section{Quenched hadron spectrum in the continuum limit}\label{sec:spectrum}
\subsection{Scale parameter}

\noindent
\begin{figure}[t]
\begin{center} \leavevmode
\epsfxsize=7.5cm \epsfbox{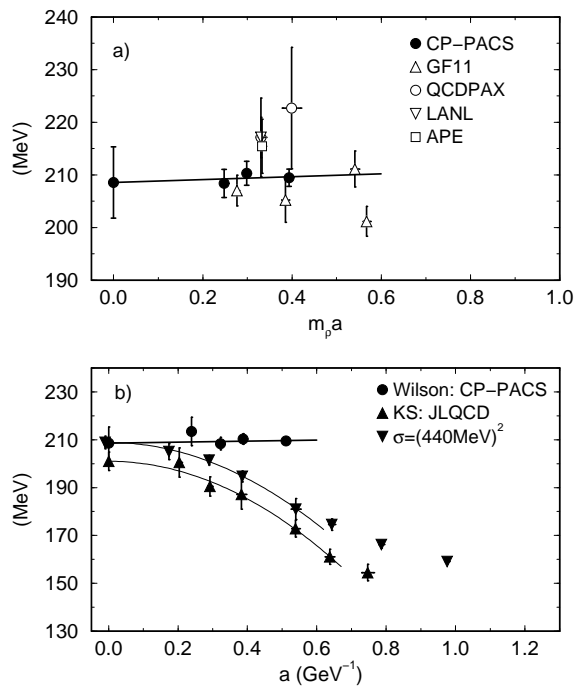}
\end{center}
\vspace{-12mm}
\caption{$\Lambda_{\overline {MS}}$ vs. $m_\rho a$ (top figure) 
and $a^{-1}$ (bottom figure).
Data for Wilson quarks are taken from 
GF11\protect\cite{ref:GF11mass},
QCDPAX\protect\cite{ref:QCDPAX96},
LANL\protect\cite{ref:LANL96}, and
APE\protect\cite{ref:APE24}.
Data for the KS action are from ref.\protect\cite{ref:JLQCD-KS}.
Data for the string tension are taken from 
ref.\protect\cite{ref:BS}.}
\label{fig:scale}
\vspace{-6mm}
\end{figure}
The scale parameter $\Lambda$ is a fundamental quantity in QCD.
We estimate the value in the $\overline{MS}$ scheme defined by
\[
\Lambda_{\overline {MS}} = 
{\pi\over a} (b_0g^2_{\overline {MS}}(\pi/a))^{-b_1/2b_0^2} 
e^{ -1/2b_0g^2_{\overline {MS}}(\pi/a)}. 
\]
where for the $\overline{MS}$ coupling we employ the 
tadpole-improved one-loop formula given by\cite{ref:LM,ref:El-Khadra} 
\[
{1/{g^2_{\overline {MS}}(\pi/a)}} = 
{ {{\rm tr}(U_P/3) / g^2} + 0.02461 }.
\]
We show the result in Fig.\ref{fig:scale}(a) as a function of $m_\rho a$. 
Our results are consistent with those reported so far
\cite{ref:QCDPAX96,ref:GF11mass,ref:LANL96,ref:APE24} 
for the Wilson action.  Our errors, however, are much reduced.
In Fig.\ref{fig:scale}(b) are compared our results
with those for the Kogut-Susskind action\cite{ref:JLQCD-KS}
and those from string tension\cite{ref:BS}.
We find that all of the results agree in the continuum limit.

\subsection{Strange mesons}

\begin{figure}[t]
\begin{center} \leavevmode
\epsfxsize=7.5cm \epsfbox{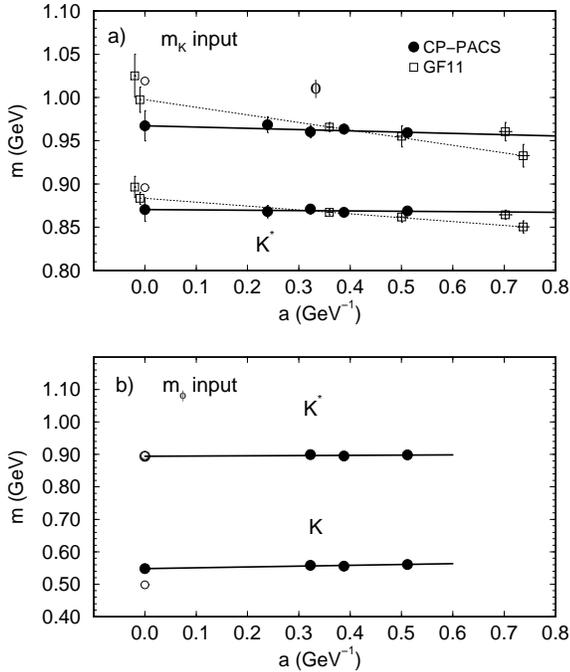}
\end{center}
\vspace{-12mm}
\caption{Strange meson masses and their extrapolations to the continuum limit.
Open circles represent experimental values.
Data from GF11\protect\cite{ref:GF11mass} are plotted by open squares 
for comparison.}
\label{fig:mesonE}
\vspace{-6mm}
\end{figure}

\begin{figure}[t]
\begin{center} \leavevmode
\epsfxsize=7.5cm \epsfbox{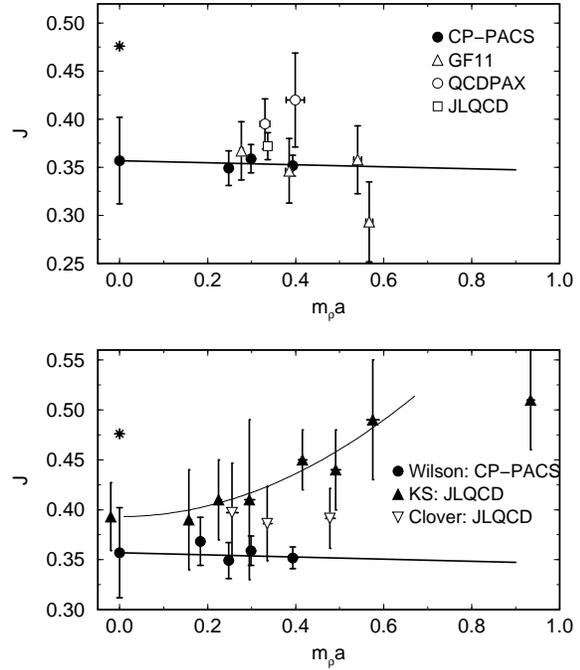}
\end{center}
\vspace{-12mm}
\caption{Value of $J$ vs. $m_\rho a$.
Data are taken from references quoted
in the caption of Fig. \protect\ref{fig:scale}
and refs. \protect\cite{ref:JLQCD1000,ref:JLQCD-Clover}.
We have calculated values of $J$ using data of 
$m_\pi$ and $m_\rho$ in the cases that values of $J$ are not
reported in the literature.}
\label{fig:J}
\vspace{-6mm}
\end{figure}

In Fig.\ref{fig:mesonE} we show the continuum extrapolation for masses of 
mesons containing a strange quark.  We find a linear behavior of the masses in 
lattice spacing to be well satisfied as shown by solid lines.

As we noted in Sec.~\ref{sec:extrap} strange quark mass can be determined 
in two ways.  In Fig.\ref{fig:mesonE}(a) we take $m_K$ as input and 
predict the masses of $\phi$ and $K^*$ meson.
The values extrapolated to the continuum disagree with experimental values 
plotted by open circles at $a=0$.

This should be contrasted with the results of the GF11 
Collaboration\cite{ref:GF11mass} plotted by open squares.
They linearly extrapolate data measured 
on lattices with a spatial extent of about 2.3 fm (dashed lines).  
They then make a finite size 
correction to the extrapolated value, and obtain a result which agree well 
with experiment. The best fit to our data differ from those by GF11 taken on 
a 2.3 fm lattice, while their data on a 3.3 fm lattice at $\beta=5.7$ 
are consistent with a linear fit of our results.

We show an alternative procedure in 
Fig.\ref{fig:mesonE}(b) where we plot results for $m_K$ and $m_{K^*}$ 
when we take $m_\phi$ as input.
In this case, the value of  $m_{K^*}$ in the continuum limit
is consistent with experiment, while that for $m_K$ is much higher.
These results mean that there are no choice of strange quark mass 
which leads to an agreement of vector as well as pseudo-scalar meson masses.

The $J$ parameter\cite{ref:J} is a convenient quantity to summarize 
the situation.
As shown in Fig.\ref{fig:J}(a), the value of $J$ in the continuum limit is
smaller than the experimental value by more than 2 standard deviations.
In Fig.\ref{fig:J}(b), we compare our results with those obtained with the  
Kogut-Susskind quark action\cite{ref:JLQCD-KS} and the clover quark 
action\cite{ref:JLQCD-Clover}.
In the continuum limit, the two actions give a result consistent  
with that of the Wilson action. 

\subsection{Baryon masses}

\begin{figure}[thb]
\begin{center} \leavevmode
\epsfxsize=7.5cm \epsfbox{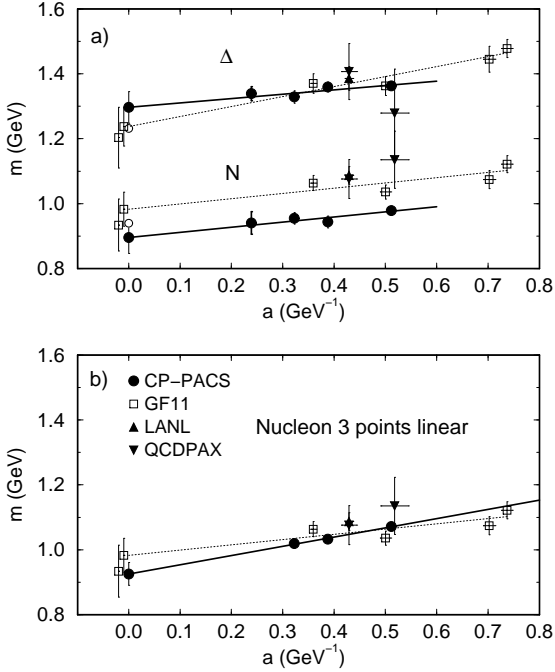}
\end{center}
\vspace{-12mm}
\caption{(a) The same as Fig.\protect\ref{fig:mesonE} for nucleon and $\Delta$.
(b) Nucleon masses obtained by linear fits to
data at the largest three quark masses.}
\label{fig:ndE}
\vspace{-6mm}
\end{figure}

Our results for nucleon and $\Delta$ masses are shown in Fig.\ref{fig:ndE}(a).
For the nucleon our continuum extrapolated value is about 5\% smaller than 
experiment.  While the difference is only a one standard deviation effect, 
this contrasts with experience of previous studies that 
the nucleon mass tends to come out higher than experiment.
In fact our results are significantly lower already at finite lattice 
spacings compared with previous results also shown in the figure.
The discrepancy is mainly due to the bending of the nucleon mass 
toward small quark masses as discussed in Sec.~\ref{sec:extrap}. 

We emphasize that our results are consistent with those from other groups 
in the region of strange quark. 
If we make a linear chiral extrapolation of the nucleon mass 
employing only three points of our data 
corresponding to $m_\pi/m_\rho \approx$ 0.75,\ 0.7, and 0.6,  
we obtain results in the chiral limit which agrees with those 
reported so far as illustrated in  Fig.\ref{fig:ndE}(b).

For $\Delta$ results from the present work and those of earlier studies 
overlap at finite lattice spacing.  
Nonetheless the continuum extrapolation leads to a value 
about 5\% larger compared to experiment in our case as compared to a value 
consistent with experiment reported by GF11.

\begin{figure}[thb]
\begin{center} \leavevmode
\epsfxsize=7.5cm \epsfbox{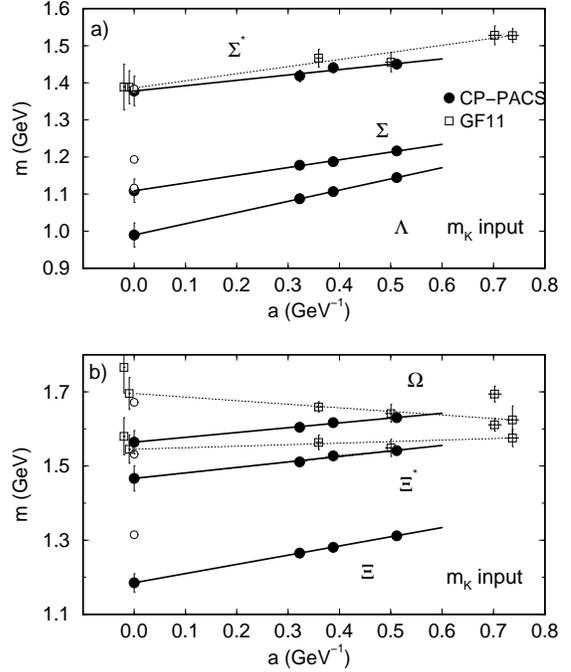}
\end{center}
\vspace{-12mm}
\caption{The same as Fig.\protect\ref{fig:mesonE} for strange baryons.}
\label{fig:strangeE}
\vspace{-6mm}
\end{figure}

Figure \ref{fig:strangeE} shows results for strange baryons employing
$m_K$ to determine the strange quark mass. 
We observe significant departures from the experiment except for $\Sigma^*$.  
Our extrapolation is quite different from that of GF11 for $\Omega$ and 
$\Xi^*$ for which GF11 found an agreement with the experiment.  

\subsection{Quenched spectrum of light hadrons}

\begin{figure}[thb]
\begin{center} \leavevmode
\epsfxsize=7.5cm \epsfbox{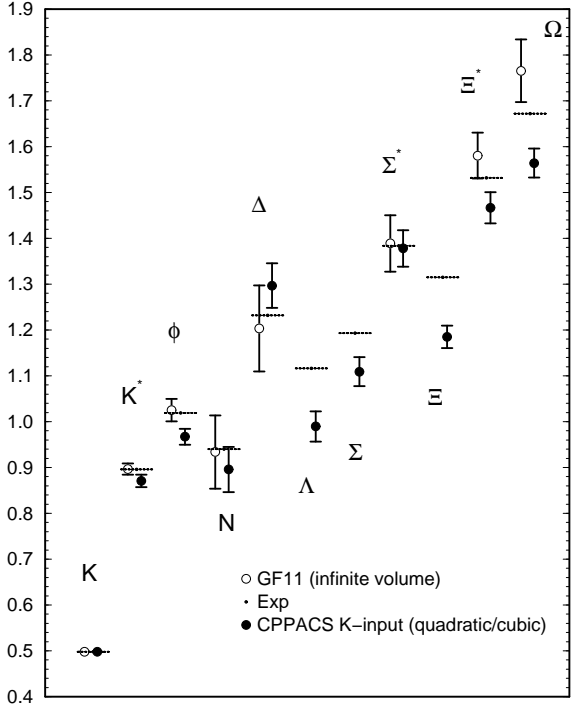}
\end{center}
\vspace{-12mm}
\caption{Results for the quenched light hadron spectrum in the continuum 
limit from the present work as compared with experimental spectrum 
(horizontal dashed bars) and those of GF11 (open circles).}
\label{fig:spectrum}
\vspace{-6mm}
\end{figure}

We show all of our results for the quenched light hadron spectrum in 
Fig.~\ref{fig:spectrum} where we also plot the experimental spectrum 
(horizontal dashed bars) and the results of GF11 (open symbols) 
for comparison.  
For strange quark mass the $K$ meson mass is used as input both in our 
results and those of GF11.  

It is quite conspicuous that the masses of baryons containing strange quark 
are systematically lower than experiment.  This may be related to a similar 
deviation in the meson sector where the masses of $K^*$ and $\phi$ are 
smaller.  

It is interesting to note that
agreement with experiment is much better if we examine the spectrum within 
each flavor $SU(3)$ multiplet. For example, 
if we take $m_\rho$ and $m_\phi$ to determine the lattice spacing
and the strange quark mass, mass of $K^*$ is consistent with experiment, and
a similar situation holds for $\Sigma^*$ and $\Xi^*$ masses 
if we take the $\Delta$ and $\Omega$ masses instead for input.
This property may contain some hint to consider the problem of
how dynamical quark effects correct the discrepancy observed in 
Fig.~\ref{fig:spectrum}.

In the up-down quark sector our value for the nucleon mass is smaller 
than experiment by about 5\% and that for $\Delta$ larger by a similar 
amount. The errors are of a similar magnitude, however.

Compared to the results of GF11, our errors are reduced by a factor two with 
our present statistics.  The spectrum itself deviate from that of GF11 
beyond estimated errors in a number of 
channels, most notable  being  $\phi$, $\Xi^*$ and $\Omega$.  

\section{Light quark masses}\label{sec:quarkmass}

\begin{figure}[t]
\begin{center} \leavevmode
\epsfxsize=7.5cm \epsfbox{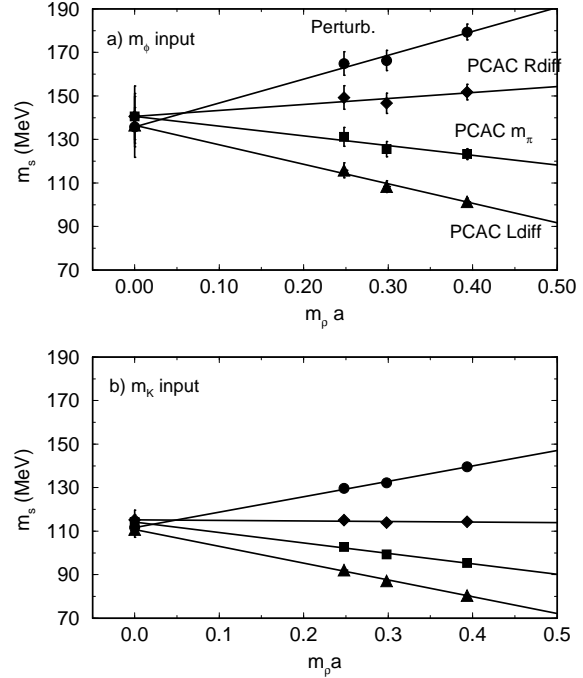}
\end{center}
\vspace{-12mm}
\caption{Strange quark masses in the ${\overline {MS}}$ scheme 
at $\mu=2$ GeV obtained from perturbation theory and
from the axial Ward identity.
Rdiff(Ldiff) stands for right- (left-) difference for an approximation
to the time derivative in the Ward identity.
Values obtained with center-difference are numerically 
close to those with $m_\pi$, and are omitted in the figure.}
\label{fig:Mstrange}
\vspace{-6mm}
\end{figure}

The values of light quark masses are important in a variety of 
phenomenological context\cite{ref:mackenzie}.  
Perturbatively quark mass is defined  by 
\begin{equation}
2m_q^{\rm Pert.} = Z_m (1/K_1+1/K_2-2/K_c)
\end{equation}
with a renormalization factor $Z_m$.  A non-perturbative 
definition is also possible based on the axial vector Ward 
identity\cite{ref:Bo,ref:Itoh86,ref:MM}
\begin{equation}
\label{eq:WI}
\langle 0| \partial_\mu A_\mu |x \rangle \ =
\ 2m_q \ \langle 0 |P |x \rangle. 
\end{equation}
We take here the definition given by
\begin{equation}
\label{eq:PCACmass}
2m_q^{\rm PCAC} = -m_\pi\ {Z_A \over Z_P} \ 
  \lim_{t\to\infty} { { \sum_{\vec x} \langle A_4({\vec x},t)\ P \rangle } 
 \over { \sum_{\vec x} \langle P({\vec x},t)\ P \rangle } },
\end{equation}
where $A_4$ is the local axial current and $P$ is the pseudo-scalar
density, and $m_\pi$ on the righthand side is an approximation to the 
time derivative in the Ward identity. 

We compare results for the two definitions of quark mass.  For the Ward 
identity method we test also $e^{m_\pi}-1$, $1-e^{-m_\pi}$, and
$(e^{m_\pi}-e^{-m_\pi})/2$ instead of $m_\pi$ for replacing the time 
derivative which correspond to right-, left- and center-difference, 
respectively.
For the renormalization constants $Z_m$\cite{ref:Zm}, 
$Z_A$ and $Z_P$\cite{ref:Zavp}, we employ one-loop results  
improved by the tadpole procedure with the coupling 
$\alpha_V(1/a)$ \cite{ref:LM}given by 
$Z_m = 8K_c (1+0.01 \alpha_V(1/a))$,
$Z_P = Z_K (1-1.034 \alpha_V(1/a))$, and
$Z_A = Z_K (1-0.316 \alpha_V(1/a))$.
Matching of lattice values of quark mass to those in the continuum 
is made at the scale $1/a$ employing the $\overline {MS}$ scheme 
in the continuum.  The result is then run by the two-loop 
renormalization group equation to 2 GeV.

\begin{figure}[t]
\begin{center} \leavevmode
\epsfxsize=7.5cm \epsfbox{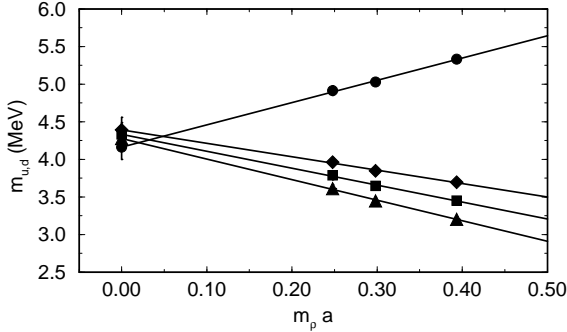}
\end{center}
\vspace{-12mm}
\caption{The same as Fig.\protect\ref{fig:Mstrange} 
for the average of up and down quark masses.}
\label{fig:Mnormal}
\vspace{-6mm}
\end{figure}

We plot results for the strange quark mass 
obtained with perturbative and three cases of non-perturbative PCAC 
definitions in Fig.\ref{fig:Mstrange}.  
The $\phi$ meson mass is used to determine $K_s$ in 
Fig.\ref{fig:Mstrange}(a) and the $K$ meson mass is employed in (b). 
For finite lattice spacings the four definitions yield significantly 
different values.  However, a linear extrapolation yields results in mutual 
agreement in the continuum limit. The same situation also holds for 
the average of up and down quark mass as shown in Fig.\ref{fig:Mnormal}.  

We note, however, that the value $m_s\approx 140$ MeV 
for the strange quark mass 
in the continuum limit obtained with $m_\phi$ as input does not agree with 
$m_s\approx 115$ MeV found with $m_K$ as input.  This is another sign of 
pathology in quenched QCD.

\begin{figure}[thb]
\begin{center} \leavevmode
\epsfxsize=7.5cm \epsfbox{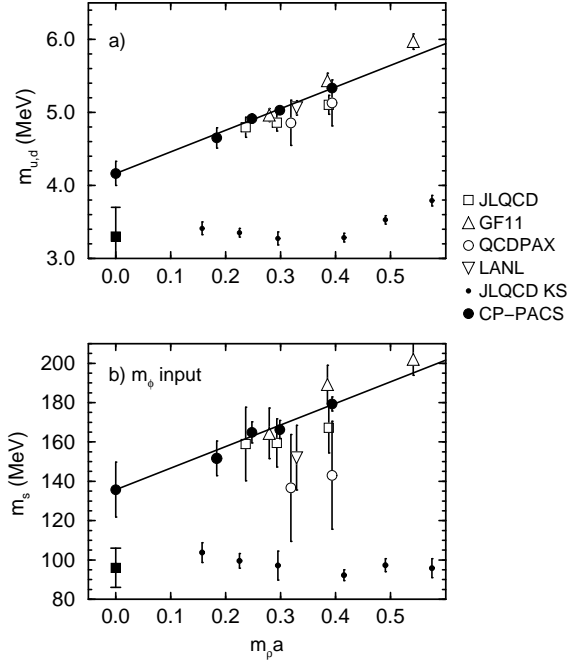}
\end{center}
\vspace{-12mm}
\caption{Values of the average of up and down quark masses (top figure)
and the strange quark mass (bottom figure).
They are determined perturbatively
in the ${\overline {MS}}$ scheme at $\mu=2$ GeV.
Data of hadron masses for Wilson quarks are 
taken from JLQCD\protect\cite{ref:HashimotoLAT96},
GF11\protect\cite{ref:GF11mass},
QCDPAX\protect\cite{ref:QCDPAX96} and
LANL\protect\cite{ref:LANL96}
to calculate quark masses by the same method
we employed for the CP-PACS data.
Results of quark masses for the KS action are
taken from JLQCD\protect\cite{ref:JLQCD-KS}.
Solid squares at $m_\rho a=0$ represent best
estimates in ref.\protect\cite{ref:LANLMQ} for the Wilson action
in the continuum limit.} 
\label{fig:mqpert}
\vspace{-6mm}
\end{figure}

In Fig.\ref{fig:mqpert}, we compare our results
for perturbative quark masses with those from other 
groups\cite{ref:HashimotoLAT96,ref:QCDPAX96,ref:GF11mass,ref:LANL96}.
Extrapolating our data to the continuum limit leads to 
$m_{u,d} \approx 4.2$ MeV  and  $m_s \approx 136$ MeV.
They are larger than the values estimated in Ref.\cite{ref:LANLMQ}
($m_{u,d} \approx 3.4$ MeV and $m_s \approx 96$ MeV) based on a compilation 
of world data for the Wilson action. In the same figure 
we also plot results for the Kogut-Susskind action obtained by the 
JLQCD collaboration\cite{ref:JLQCD-KS}.
It is not clear if Wilson and Kogut-Susskind results agree 
in the continuum limit.

\section{Decay constants}\label{sec:decay}

Determination of hadronic matrix elements often suffers from 
uncertainties in renormalization factors.  
An exceptional case is the vector meson decay constant $f_V$
for which the renormalization constant for the local vector current
$V^L$ can be precisely determined non-perturbatively 
with the use of the conserved current $V^C$ \cite{ref:MM}.
It has been noted by the QCDPAX Collaboration that
this determination of the renormalization factor leads to a
remarkable scaling of the decay constants\cite{ref:QCDPAX96}.

\begin{figure}[thb]
\begin{center} \leavevmode
\epsfxsize=7.5cm \epsfbox{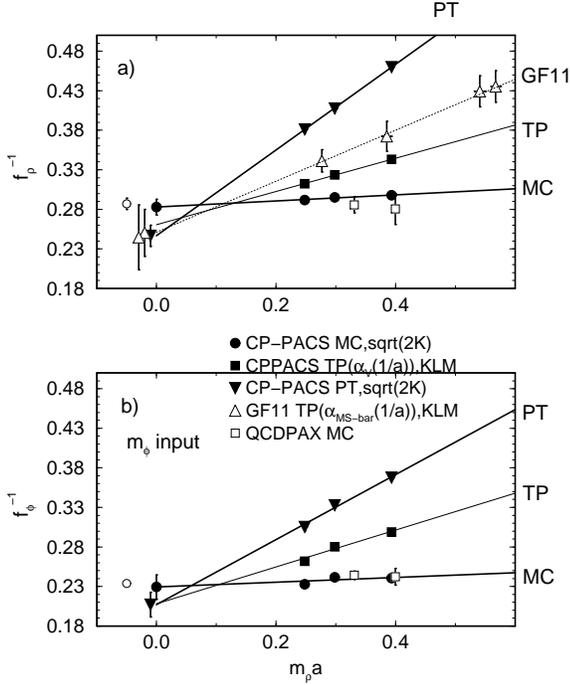}
\end{center}
\vspace{-12mm}
\caption{Results of $f_\rho$ and $f_\phi$ obtained 
with various renormalization constants.
GF11 results \protect\cite{ref:GF11decay} are plotted for comparison.}
\label{fig:fV}
\vspace{-6mm}
\end{figure}

In Fig.\ref{fig:fV} we plot $f_V$ obtained with the renormalization 
factor calculated non-perturbatively through
$Z_V=2K {\tilde Z_V}$ with ${\tilde Z_V}$
determined from ${\langle V^C|V^L \rangle}/{\langle V^L|V^L \rangle}$. 
Results obtained with naive perturbative $Z$ factor to one loop given by  
$Z_V = 2K\ (1 - 2.19\ \alpha_{latt.})$\cite{ref:Zavp} 
and with tadpole-improved perturbation theory
$Z_V = (1-3K/4K_c)\ (1 - 0.82\ \alpha_V(1/a))$
\cite{ref:LM,ref:ZKtp} are also shown for comparison.

Making a linear extrapolation of the non-perturbative result 
in $m_\rho a$, we obtain a value in the continuum limit which is
completely consistent with experiment.
Perturbative renormalization factors lead to slightly
smaller values. It is worth noting that non-perturbative 
renormalization constant has a small value in the range of our data, 
given by ${\tilde Z_V} \approx$ 0.54,\ 0.60, and 0.64 at
$\beta=$ 5.9,\ 6.1,\ and 6.25, respectively.
A reliable extrapolation of the decay constants with
one-loop evaluation of renormalization factors may be difficult in such 
a situation.

\begin{figure}[t]
\begin{center} \leavevmode
\epsfxsize=7.5cm \epsfbox{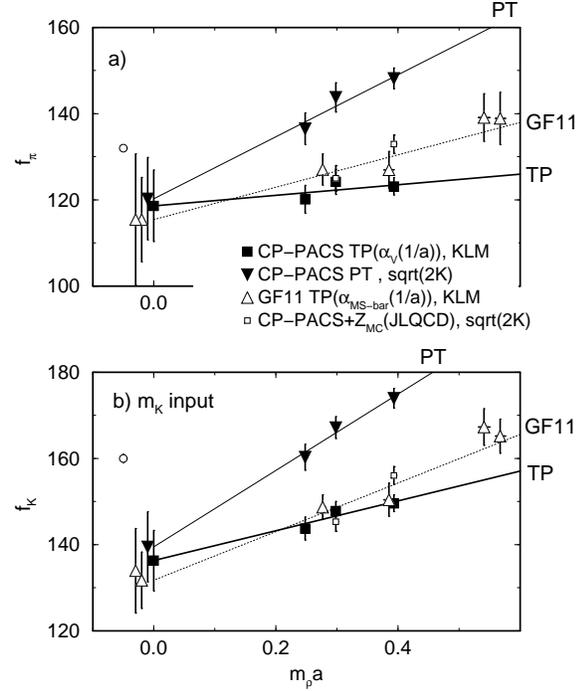}
\end{center}
\vspace{-12mm}
\caption{The same as Fig.\protect\ref{fig:fV} for
pseudo-scalar meson decay constants $f_\pi$ and $f_K$.}
\label{fig:fP}
\vspace{-6mm}
\end{figure}

For pseudo-scalar meson decay constant $f_P$, the naive one-loop 
perturbative $Z$ factor takes the form 
$Z_A = 2K\ (1 - 1.68\ \alpha_{latt.})$ \cite{ref:Zavp}
and a tadpole-improved value is given by 
$Z_A = (1-3K/4K_c)\ (1 - 0.316\ \alpha_V(1/a))$ \cite{ref:LM,ref:ZKtp}.
Both of these renormalization constants lead to a consistent value
in the continuum limit as shown in Fig.\ref{fig:fP}. 
Values of $f_\pi$ and $f_K$ in the continuum limit
are smaller than the experimental values by one to two standard deviations.

A non-perturabtive estimate of the axial vector renormalization factor is 
available at $\beta=5.9$ and 6.1\cite{ref:KuramashiLAT96,ref:YoshieLAT96}.
Extrapolations of our data with these results for non-perturbative $Z_A$
seems to give a consistent value with that with perturbative $Z_A$.

\section{Conclusions}
In this article we have presented a status report of our effort toward 
precision results of light hadron spectrum in quenched QCD with the 
standard plaquette and Wilson quark actions. 

Our results have confirmed that strange quark mass cannot be tuned in 
quenched QCD so that both pseudo scalar and vector spectra are in agreement 
with experiment.  This observation is extended to the strange baryon 
sector in the present study, 
where our values for the baryon masses are systematically lower than 
experiment.

We have also found that the nucleon mass decreases faster than a linear 
behavior in quark mass toward the chiral limit.  
As a result the continuum value is  smaller than the experiment, 
contrary to results of previous studies, albeit the difference is still a 
one standard deviation effect.  

Finally, we have shown that
quark mass determined from the axial Ward identity agrees with
that from perturbation theory in the continuum limit, providing a handle 
for controlling the continuum extrapolation. 

In parallel with these points of physics, an important lesson we would like
to draw from the present work is the importance of precision; with errors 
down to a percent level uncertainties in physics implications can be 
much removed.

Encouraged by the results, we plan to further increase statistics of the three 
runs at $\beta=$5.9, 6.1 and 6.25.  In addition 
we shall soon have results from the run at $\beta=6.47$.
We hope that results from the additional simulations  help solidify our 
findings so far, and lead to a standard of the quenched hadron spectrum in QCD.

\section*{Acknowledgements}

This work is supported in part by the Grant-in-Aid
of Ministry of Education, Science and Culture
(Nos.\ 08NP0101, 08640349, 08640350, 08640404, 08740189,
and 08740221).  Two of us (GB and RB) are supported by
the Japan Society for the Promotion of Science.

\end{document}